\documentclass{aa} 
\usepackage{natbib}
 \usepackage{color}
     \usepackage{xcolor}
\bibpunct{(}{)}{;}{a}{}{,}

\usepackage{graphicx}

\begin{document}

   \title{Convective mixing in distant and close-in giant planets}
   
   \subtitle{Dependences on the initial composition, luminosity, bloating, and semi-convection}

   \author{J. Polman
          \and
          C. Mordasini
          }

   \institute{Division of Space Research and Planetary Sciences, Physics Institute, University of Bern, Gesellschaftsstrasse 6, 3012 Bern, Switzerland}
             
   \date{Received August 15, 2024 / Accepted November 22, 2024}

\abstract{Recent structure models of Jupiter suggest the existence of an extended region in the deep interior with a high heavy element abundance, referred to as a dilute core. This finding has led to increased interest in modelling the formation and evolution processes with the goal of understanding how and under what circumstances such a structure is formed and retained, to in turn better understand the relation between atmospheric and bulk metallicity. We modelled the evolution of giant planets, varying various parameters relevant for the convective mixing process, such as the mixing length parameter and the size of the mesh, and parameters related to the general evolution, such as the orbital distance and the initial luminosity. We in particular studied hot Jupiters and find that the effect of bloating on the mixing process is small but can in some cases inhibit convective mixing by lowering the intrinsic luminosity for a given entropy. Semi-convection can significantly lower the extent of a dilute core if it is strong enough. We find that dilute cores are unable to persist for initial luminosities much higher than $\sim 3 \times 10^3$ L$_J$ for a Jupiter-like planet for the initial heavy element profiles we studied. From this we conclude that, based on our model, it is unlikely that a large number of giant planets retain a dilute core throughout their evolution, although this is dependent on the assumptions and limitations of our method. Future work should focus on improving the link between formation and evolution models so that the mixing process is accurately modelled throughout a planet's lifetime and on improving the understanding of how to model convection near radiative-convective boundaries.}

\maketitle
%

\section{Introduction}
In recent years vast amounts of atmospheric data from exoplanets have been gathered using telescopes such as \textit{Hubble}, \textit{Spitzer,} and various ground-based telescopes \citep{Madhusudhan2019_ReviewAtmospheres}. JWST makes it possible to characterise atmospheres with a quality previously unimaginable, helping us constrain elemental abundances within these atmospheres \citep{GuzmanMesa+2020_JWSTNeptunes,Alderson+2023_WASP39b,Dyrek+2024_WASP107b,Benneke+2024_TOI270d}. Work has also been done to constrain the bulk heavy element mass of giant planets using mass and radius data \citep{Thorngren2016_BulkMetalMass}, although such efforts are often complicated due to radius inflation \citep{Demory2011_Inflation,Sarkis+2021_Bloating}. Since the early work of \citet{Oeberg2011_COratios}, efforts have also been made to track elemental abundances through the formation process \citep{Mordasini2016_Formation,Turrini2021_TracingFormationHistory,Khorshid2022_SimAb}.

In such studies it was often assumed that atmospheric characteristics could be related to bulk properties and the accretion of different chemical elements in a relatively simple manner, under the assumption that giant planets consist of a well-defined core and a convective homogeneous envelope \citep{Guillot2005_ReviewCoreEnvelope}, although the idea of Jupiter having a more complex compositional structure had been suggested \citep{Stevenson1982_JupiterReview,Vazan+2015_Convection}. Recent measurements by the Juno mission \citep{Bolton2017_Juno} suggest the presence of a dilute core in Jupiter's interior \citep{Wahl2017_DiluteCoreJupiter,Debras+2019_Jupiter,Miguel2022_Jupiter}, raising questions about how such dilute cores form and whether the simplified assumption remains valid for exoplanets. This inspired work on how Jupiter's dilute core formed and was able to persist throughout its evolution \citep{Vazan+2018_Jupiter,Mueller+2020_Jupiter,Stevenson+2022_formation}. These measurements also raise the question of whether we should expect dilute cores to be present in exoplanets.

Recent work by \citet{Vazan+2018_Jupiter} focused on replicating Jupiter's structure based on the observational constraints found by Juno, which they did by identifying the parameters at the start of evolution to find fitting models. \citet{Mueller+2020_Jupiter} tracked planets throughout the formation and evolution process using several different formation models and varied several of the parameters relevant for retaining a dilute core. \citet{Stevenson+2022_formation} tracked the formation and evolution of a planet across each stage in depth. They discuss some of the effects a dilute core has on the evolution, but focus less on the mixing process.

We modelled the evolution of various giant planets, incorporating convective mixing. We analysed the role different parameters play and discuss their role in determining whether dilute cores can be retained. We incorporated additional internal heating to accurately model close-in planets and discuss the role dilute cores should play in interpreting atmospheric data. This paper is structured as follows. In Sect. \ref{Sec:Method} we describe the evolution model and explain the initial parameters. In Sect. \ref{Sec:Results} we present our results for the standard models, as well as the impact the values of several parameters have on the final composition. In Sect. \ref{Sec:Limits} we discuss some of the limitations of our current model and the effect this has on our results. In Sect. \ref{Sec:retaincore} we analyse our findings in depth and discuss the necessary conditions for retaining a dilute core. We summarise our findings in Sect. \ref{Sec:summary}.

\section{Method} \label{Sec:Method}
\subsection{Evolution model} \label{Sec:Evo_model}
We used the planetary evolution model {\tt completo21} \citep{Mordasini+2012_Completo} as the framework for our study but improved it in several ways. The internal structure of gas giants was computed in 1D using Eqs. \ref{eq:dmdr}-\ref{eq:dTdr}: 
\begin{equation} \label{eq:dmdr}
    \frac{dm}{dr}=4\pi r^2 \rho
\end{equation}
\begin{equation} \label{eq:dPdr}
    \frac{dP}{dr}=\frac{-Gm}{r^2}\rho
\end{equation}
\begin{equation} \label{eq:dldr}
    \frac{dl}{dr}=4\pi r^2 \rho \frac{L}{M}
\end{equation}
\begin{equation} \label{eq:dTdr}
    \frac{dT}{dr}=\frac{T}{P}\frac{dP}{dr}\nabla (T,P)
.\end{equation}
In Eq. \ref{eq:dTdr}, $\nabla (T,P)$ follows from the Ledoux criterion
\begin{equation}
    \nabla_{rad}<\nabla_{ad}+\nabla_X,
\end{equation}
while previous versions of the model used the simple Schwarzschild criterion, which gives $\nabla (T,P)$=min$(\nabla_{rad},\nabla_{ad})$ assuming a homogeneous composition in the envelope. Here,
\begin{equation} \label{eq:radGrad}
    \nabla_{rad}=\frac{3P\kappa L}{16\pi acGmT^4}
\end{equation}
is the radiative-conductive gradient, with the opacity given by
\begin{equation}
\frac{1}{\kappa}=\frac{1}{\kappa_{rad}}+\frac{1}{\kappa_{cond}},
\end{equation}
where the radiative opacity follows from the opacities of \citet{Bell+1994_radopacity}, which are independent of the metallicity, where we assume that the atmosphere is grain/condensate free. The conductive opacity follows from the opacity tables of \citet{Cassisi2007_condopacities}, which are dependent on $\bar{Z}$. It is important to note that $\nabla_{rad}$ thus refers to both the radiative and conductive contribution, with the conductive contribution generally being dominant in the interior and the radiative contribution only being dominant in the atmosphere. The remaining quantities $\nabla_{ad}$ and 
\begin{equation} \label{eq:nablaX}
\nabla_X=\frac{\partial \text{ln} T(\rho, p, X)}{\partial X_j}\frac{d X_j}{d \text{ln} p}
\end{equation}
are the adiabatic gradient and the compositional gradient, respectively, where X$_j$ is the mass fraction of the jth species and we sum over j. When the stability criterion is satisfied the area is radiative-conductive. When the criterion is not satisfied the area is convective, with efficient convection leading to $\nabla (T,P)\approx\nabla_{ad}+\nabla_X$, while inefficient mixing results in $\nabla (T,P)\approx\nabla_{rad}$. A detailed derivation of $\nabla (T,P)$ can be found in the appendix of \citet{Vazan+2015_Convection}. At low pressures we reverted to the Schwarzschild criterion since the compositional gradient at these pressures is governed by atmospheric effects not considered in this work. We incorporated the double-grey atmosphere model of \citet{Guillot2010_grey} to model the impact of stellar irradiation on the outer boundary condition. We used the equations of state of \citet{Chabrier+2021_HHeEOS} for a mixture of hydrogen and helium and the equations of state of \citet{Haldemann+2020_H2OEOS} for water. In this work water was used to represent the general high-Z material. 
We changed Eq. \ref{eq:dldr} from $\frac{dl}{dr}=0$ (as in our earlier models) to an interior luminosity that scales with the mass to better approximate the luminosity interior of the planet as well as the radiative-conductive gradient. The total luminosity, L, was still determined using the method described in \citet{Mordasini+2012_Completo}. The effect of this approximation and alternative methods are discussed in Sect. \ref{sec:luminosity_simp}.

Convective mixing is implemented on a mesh in mass-space using the Crank-Nicolson approximation. The mixing velocity at each mesh point follows from the Ledoux criterion combined with an estimate by \citet{Mihalas1978_convection} for the efficiency of energy transport by convection. The method was implemented as described in detail in the appendix of \citet{Vazan+2015_Convection}. The diffusion coefficient is given by 
\begin{equation}
    D=0.1vl=0.1(l/H_p)vH_p,
\end{equation}
where the factor 0.1 gives the efficiency of convection, in line with previous studies that used the same value \citep{Vazan+2015_Convection,Vazan+2018_Jupiter}. $H_p$ is the pressure scale height and $l/H_p=\alpha$ is the mixing length parameter giving the mean free path of convective elements. We used a value of 10$^{-3}$ for $\alpha$ unless otherwise specified. The change of metallicity at each point is given by 
\begin{equation}
    \frac{dZ}{dt}=-\frac{\partial F}{\partial m},    F=-(4\pi r^2\rho)^2 D \frac{\partial Z}{\partial m},
\end{equation}
where F is the particle flux. Semi-convection can occur when 
\begin{equation}
    \nabla_{ad}<\nabla_{rad}<\nabla_{ad}+\nabla_X,
\end{equation}
which is when convection is inhibited in the Ledoux criterion, but not in the Schwarzschild criterion. We modelled semi-convection using the approximation by \citet{Langer+1983_semicon1,Langer+1985_semicon2}, where the diffusion coefficient is given by

\begin{equation}
    D=\frac{\alpha_{sc}K}{6C_p\rho}\frac{\nabla_{rad}-\nabla_{ad}}{\nabla_{ad}+\nabla_X-\nabla_{rad}}, K=\frac{4acT^3}{3\kappa\rho}
.\end{equation}

When semi-convection occurs, $\nabla$(T,P)=$\nabla_{rad}/(1+\frac{L_r^{sc}}{L_r^{rad}}$), where L$_r^{sc}$ is the semi-convective luminosity given by Eq. 13 of \citet{Langer+1983_semicon1}.
In our standard model the efficiency of semi-convection $\alpha_{sc}$ was set to 0, which stops semi-convection completely. Instead, we study the effect of semi-convection in Sect. \ref{sec:semi-con}.

Following the approach in \citet{Kippenhahn2012_MLTequations}, and also to keep the runtime of the code manageable, the thermodynamic evolution of the planet and the convective mixing have been separated numerically. The general cooling and contraction is modelled as normal with the updated structure equations. After an initial structure is found the convective mixing is computed on a separated mesh for the duration of the general evolution timestep using smaller mixing timesteps limited by the maximum allowed change in metallicity at each mesh point of 10$^{-4}$. The structure is then recomputed with the new compositional structure while taking the general evolution into account. Convective mixing is then again computed over the new timestep using the metallicity based on the mesh of the last timestep. The metallicity of the mesh is stored between steps to prevent numerical mixing due to repeatedly switching between the evolutionary structure and the mesh. The size of the thermodynamic evolution timestep is limited to a fraction of the Kelvin-Helmholtz timescale and starting at a smaller size initially. For typical conditions while mixing occurs the thermodynamic evolution timestep contains $\sim$10$^4$ mixing timesteps. The size of the mixing timestep was first estimated based on the mixing velocity and the size of the compositional gradients, with the timestep being decreased if the change in metallicity exceeds the 10$^{-4}$ limit anywhere in the structure. The thermodynamic evolution timestep and the convective mixing timestep increase as the planet cools down and the planet moves towards a stable compositional structure. We find that using smaller evolution or mixing timesteps only has a minimal effect on the final compositional structure and the general planetary properties. The interpretation of these small differences is discussed in Sect. \ref{sec:steps}. The mesh is constructed in equal steps of mass and consists of $5 \times 10^4$ points. The effect of the number of mesh points on the mixing process and its role in determining the final structure are discussed in Sect. \ref{sec:gridpoints}.

To analyse the effect of convective mixing on a planet during its evolution, we incorporated the initial compositional structures of \citet{Vazan+2018_Jupiter} and \citet{Mueller+2020_Jupiter}. We chose to incorporate these compositional structures as given, but we slowly increased the radiative-conductive gradient for mixing purposes only over 1 Myr in our model, given that we did not model the coupled formation and evolution phases. This was done to prevent instantaneous mixing at the start of the simulations, leaving the possibility for larger compositional gradients to form and merge. Instantaneous mixing was no problem in \citet{Vazan+2018_Jupiter}, most likely due to the very low initial luminosity of $\sim25$ L$_J$, while for \citet{Mueller+2020_Jupiter} convection was largely inhibited at early times due to the deposition of the accretion's shock energy creating a large entropy gradient and radiative-conductive zone in the envelope. This mainly applies for their hot-start planets, while for cold-start planets the lack of instantaneous mixing can be explained by the lower luminosity and the presence of large compositional gradients. We analyse and discuss the effects of and the justifications for delaying convective mixing in Sect. \ref{sec:delayedconv}. As our base case, we used Jupiter, with an initial luminosity of 10$^3$ L$_J$, similar to the value used for the cold-start scenarios of \citet{Mueller+2020_Jupiter}. The value we used is significantly lower than the cold accretion post-formation luminosities of $\sim2 \times 10^4$ L$_J$ to $\sim6 \times 10^4$ L$_J$ found by \citet{Mordasini2013_Luminosity} for a 1 M$_J$ planet. We note that we used a value for L$_J$ of 11.976 $\times$  10$^{-10}$ L$_\odot$, based on the results of \citet{Li+2018JupiterLum}, which is higher than the values of L$_J$ used in previous studies. We assumed a value for Y of 0.274 for Z=0.0153, in agreement with protosolar values \citep{Lodders2009_protosolar}. The ratio of helium to hydrogen was kept constant, independent of the metallicity or location in the structure. Our simulations ran for 4.5 billion years. We find that the exact runtime is insignificant in determining the final structure, due to almost all of the mixing happening in the first 10$^9$ year after formation \citep{Vazan+2018_Jupiter,Mueller+2020_Jupiter}.

\subsection{Additional internal heating for hot Jupiters}
\begin{figure}
    \resizebox{\hsize}{!}{\includegraphics{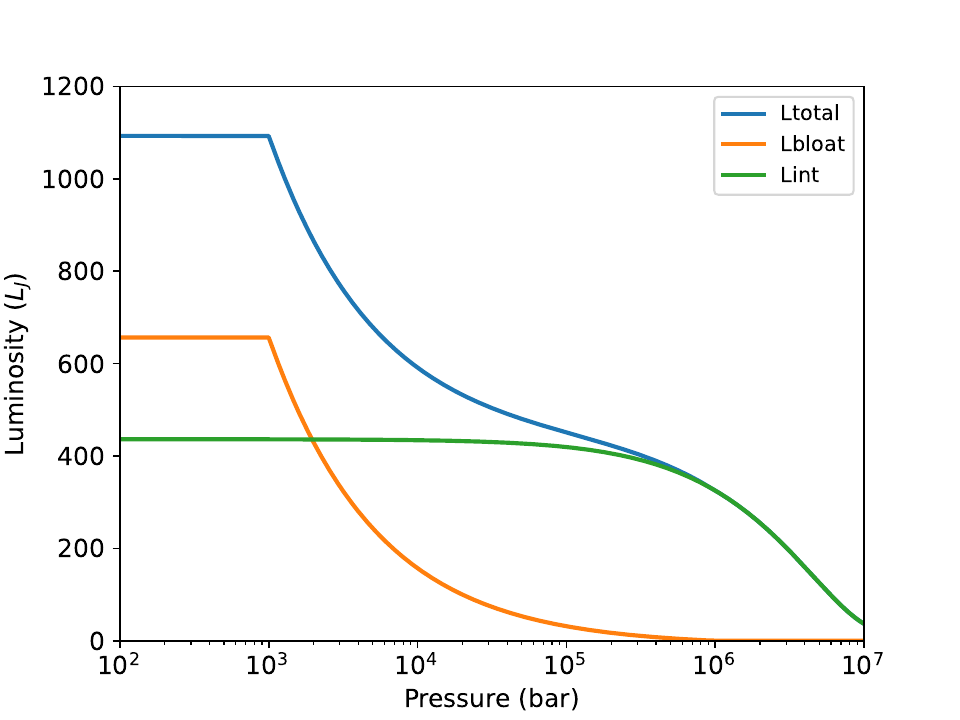}}
        \caption{Total luminosity (blue), bloating luminosity (orange), and intrinsic luminosity (green) as a function of pressure for a planet at 0.03 AU. The intrinsic luminosity scales as $\frac{dl}{dm}=\frac{L}{M}$, and the bloating luminosity follows a power law, $L \propto P^{-0.6}$, between 10$^3$ and 10$^6$ bar.}
    \label{Fig:Bloating_luminosity}
\end{figure}

We incorporated an additional internal heating term to test the relevance of the bloating process to convective mixing for hot Jupiters. Hot Jupiters are interesting objects of study, due to their size. The large radii of hot Jupiters, a result of the bloating due to the small orbital distance, makes them ideal candidates for atmospheric characterisation using transmission spectroscopy \citep{Alderson+2023_WASP39b}. Planets at small orbital distances are also more likely to transit, due to which a significant fraction of known transiting exoplanets experience this additional internal heating. We implemented bloating by depositing a portion of the incoming stellar flux in the planet. We no longer used the $\frac{dl}{dr}=0$ simplification at this point because we did not use fully convective structures where the luminosity profile is not important. Because of this we could not simply deposit the bloating luminosity in the core. Instead, we deposited the heating in the envelope separately from the internal luminosity from cooling and contraction. We deposited the energy at pressures above 10$^3$ bar with a cutoff at 10$^6$ bar. We followed a power law of $L \propto P^{-0.6}$, in line with the relation derived by \citet{Ginzburg2016_OhmicHeating} for Ohmic heating. The effect of this choice is discussed with the results in Sect. \ref{Sec:BloatResults}. To determine the efficiency of the stellar flux deposition, we compared our model to the current-day radius of HD 209458 b. We find that an efficiency of 0.5\% is best able to reproduce this radius, as well as radius inflation in general. This is lower than the 1\% found by \citet{Komacek2020_ReinflationParameterStudy} for a deposition depth of 10$^3$ bar, but not unreasonable due to the large uncertainties in determining heating efficiencies. \citet{Sarkis+2021_Bloating} also find that a heating efficiency of 0.5\% could apply for HD 209458 b. In cases where internal heating is relevant, we did not use an initial luminosity of 10$^3$ L$_J$ but instead a luminosity for which the average entropy is equivalent to that of the planet without internal heating at the same initial luminosity, which is dependent on the initial composition. An example of the luminosity is shown in Fig. \ref{Fig:Bloating_luminosity},  where at high pressures the internal luminosity determines the gradient, with the bloating becoming relevant at lower pressures. At very low pressures the heating was not implemented, but this should not have a large effect, since it is expected that compositional gradients cannot persist so far out \citep{Helled+2024_ReviewSolar}. 

\subsection{Initial metallicity distribution}
We ran our models with a core of 1 M$_E$ consisting of Z=1 material. This was done mainly to ensure convergence for each run, which would not be guaranteed with a lower core mass. For the initial envelope metallicity distribution, we used structures based on the post-formation composition found by \citet{Mueller+2020_Jupiter} and the initial composition used by \citet{Vazan+2018_Jupiter}. We did not include the Hot\_extended\_Z initial structure from \citet{Mueller+2020_Jupiter} due to its similarity to the Cold\_extended\_Z initial structure and the fact that all our models start at the same initial luminosity, removing the main difference between them. The initial compositions are presented in Fig. \ref{Fig:Overview_standardmodel}, and a summary of the standard parameters we used can be found in Table \ref{tab:paraCompleto}.

\begin{table}[]
\caption{Standard parameters used in {\tt completo21} for the Jupiter benchmark calculations.}
\label{tab:paraCompleto}
\centering
\begin{tabular}{|c|c|}
\hline
L$_{init}$            & 10$^3$ L$_J$                   \\ \hline
M$_{core}$            & 1 M$_E$                      \\ \hline
M$_{envelope}$             & 316.83 M$_E$                    \\ \hline
Y             & 0.274                     \\ \hline
$\alpha$         & 10$^{-3}$    \\ \hline
$\alpha_{sc}$          & 0                         \\ \hline
n$_{mesh}$            & 5 $\times$ 10$^4$ \\ \hline
Orbital distance & 5.2 AU                    \\ \hline
M$_\star$            & 1 M$_\odot$                    \\ \hline
\end{tabular}
\end{table}

\section{Results} \label{Sec:Results}
In this section we present the final compositional structures for different initial conditions and evolution parameters. First we present the results for our Jupiter-like planet using the parameters shown in Table \ref{tab:paraCompleto}. We then vary the values of some of the parameters to determine the role these parameters play in forming and retaining dilute cores, including in particular close-in planets. 

\subsection{Standard model: Jupiter comparison case}
\begin{figure}
    \resizebox{\hsize}{!}{\includegraphics{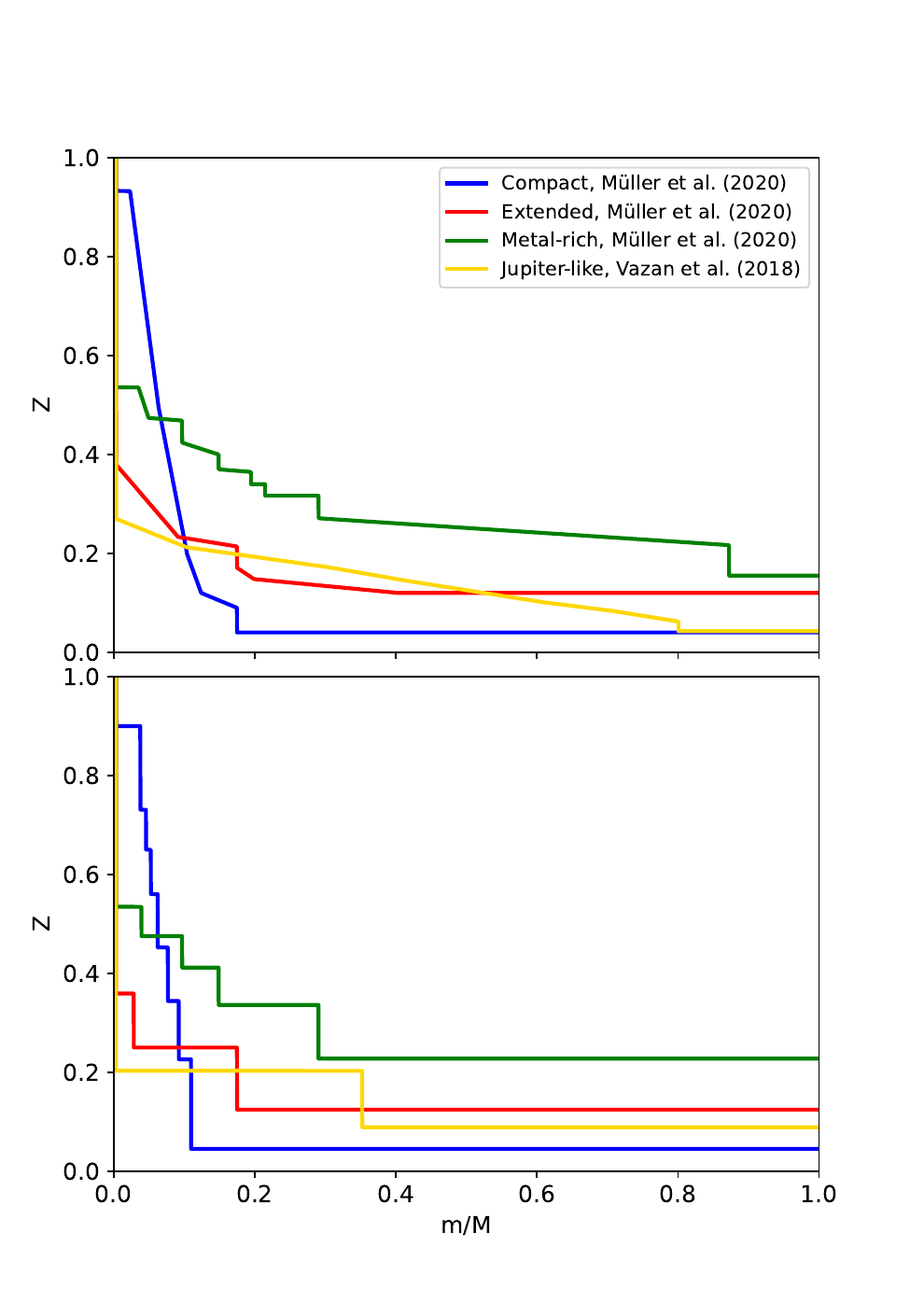}}
        \caption{Composition of the different models at the start (top) and end (bottom) of the simulation at 4.5 Gyr. The x-axis shows the normalised mass and the y-axis the fraction of high-Z material, water in our case. The compositions `compact', `extended', and `metal-rich' correspond to Hot\_Compact\_Z, Cold\_extended\_Z, and Cold\_high\_Z in \citet{Mueller+2020_Jupiter}, respectively.}
    \label{Fig:Overview_standardmodel}
\end{figure}

The results of our standard model for a Jupiter-like planet can be seen in Fig. \ref{Fig:Overview_standardmodel}, which shows the initial and final compositional structure using the standard parameters for our four different initial structures. All initial compositional structures retain a dilute core, with the one based on the structure used by \citet{Vazan+2018_Jupiter} retaining the simplest structure with only one step. This seems to be due to the small compositional gradients present in the structure and the relatively small difference between the metallicity at the bottom and the top of the envelope.

The other structures retain a more complex dilute core, with the compact model forming a structure with seven steps, the extended model forming two steps, and the metal-rich model forming four steps. We find that the creation of these steps is necessary to prevent convective mixing, as they lead to large compositional gradients, which cannot be achieved by a more gradual slope. The occurrence of steps was first reported by \citet{Vazan+2015_Convection} and is also seen in \citet{Vazan+2018_Jupiter} and \citet{Mueller+2020_Jupiter}. We find that none of the dilute cores extend beyond 35\% of the mass, with the Jupiter-like initial composition extending out the furthest. There are two reasons why it becomes harder for dilute cores to extend to a higher percentage of the mass. To have the dilute core extend to higher masses would require more high-Z material in the interior to be able to form a step of the same size. The radiative-conductive gradient also continues to increase further from the core, requiring even larger step sizes to prevent convection. The dilute core extending to 35\% of the mass differs from the results of \citet{Mueller+2020_Jupiter} and \citet{Stevenson+2022_formation}, who both find that the outer 80\% of the mass is fully mixed. These results are more in line with what we found for the compact and extended initial compositions, which both extend to less than 20\% of the mass. \citet{Vazan+2018_Jupiter} find a dilute core extending further out, but this can again be explained via the low initial luminosity they use. The extent of the dilute core for the extended, metal-rich and Jupiter-like models also agree with \citet{Howard2023_DiluteCoreExtend}, who predict that Jupiter's dilute core extends to between 15 and 60\% of its total mass based on data from Juno, while the dilute core extends to $\sim$10\% of the total mass for our compact model. Recently, \citet{Knierim+2024_ConvectiveMixing} presented models with dilute cores extending to between 5 and 30\% of the total mass for a Jupiter-mass planet, although a large core of 64 M$_E$ is necessary to retain such large dilute cores. \citet{Knierim+2024_ConvectiveMixing} also find that steep compositional gradients, like our compact case, are more likely to retain dilute cores than shallow compositional gradients, like our Jupiter-like case, which can also be seen in our results in the number of steps that can be formed. This relation becomes clearer in the remainder of this section, where the values of various parameters are altered.

The internal temperatures we find are significantly higher than those found in the paradigm where interiors are fully mixed. For the compact, extended, metal-rich, and Jupiter-like initial compositions, we find initial internal temperatures of $1.8 \times 10^5$ K, $7.3 \times 10^4$ K, $9.0 \times 10^4$ K, and $6.1 \times 10^4$ K, respectively, at the core-envelope boundary. These values are higher than the $5.0 \times 10^4$ K found by \citet{Mueller+2020_Jupiter}, especially for the compact structure, although they mention having hotter inner envelopes in some cases. The causes of discrepancies with previous works and the effect of the high internal temperatures are discussed in Sect. \ref{sec:hightemps}.

\begin{figure}
    \resizebox{\hsize}{!}{\includegraphics{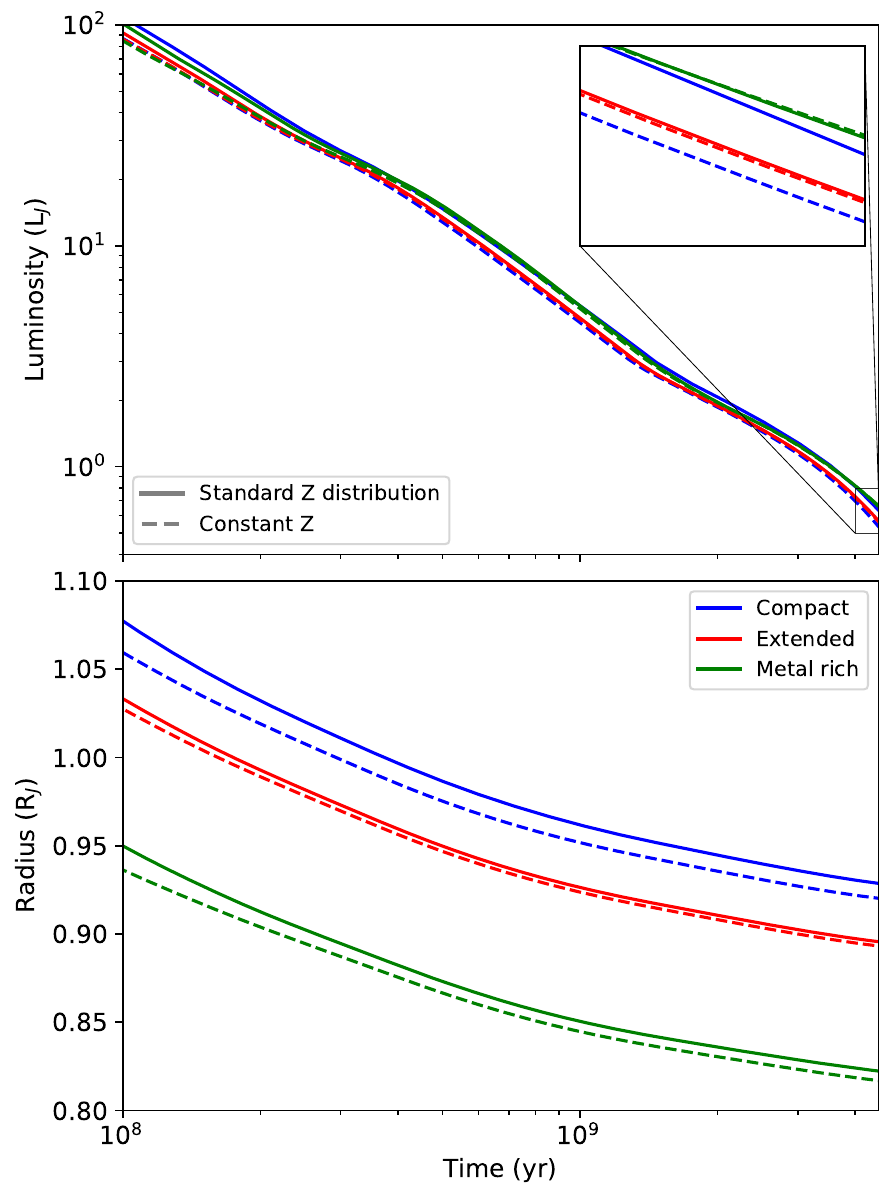}}
        \caption{Luminosity (top) and radius (bottom) for the models of our standard case and when changing the initial composition to consist of a fully mixed envelope with the same average Z as the models for the standard case, which are Z=0.103, 0.149, and, 0.278 for the compact, extended, and metal-rich initial compositional structures, respectively. The inset shows a closer look at the luminosity near the end of the evolution, which ranges from 0.54 to 0.67 L$_J$.}
    \label{Fig:LumiandRad}
\end{figure}

To determine the effect of the mixing process on the evolution of the planet, we compared the luminosity and radius for these models to those with a fully mixed envelope. This is shown in Fig. \ref{Fig:LumiandRad}. We find slightly higher present-day luminosities and radii for the new models compared to those with a fully mixed envelope. The differences in the radii are small compared to the effect of the average of Z, while for the luminosity the effect of the average of Z and of whether the envelope was mixed or not had a similar effect. The mixing process leads to an increase in the luminosity early on, but there is also a difference in the way the luminosity evolves. The increase in the luminosity due to the mixing is further explained in Sect. \ref{sec:InitLum}. Overall the effect of the mixing process on the luminosity and radius is rather small compared to other factors, at least in these cases.

\subsection{Additional internal heating: A bloated hot Jupiter} \label{Sec:BloatResults}
\begin{figure}
    \resizebox{\hsize}{!}{\includegraphics{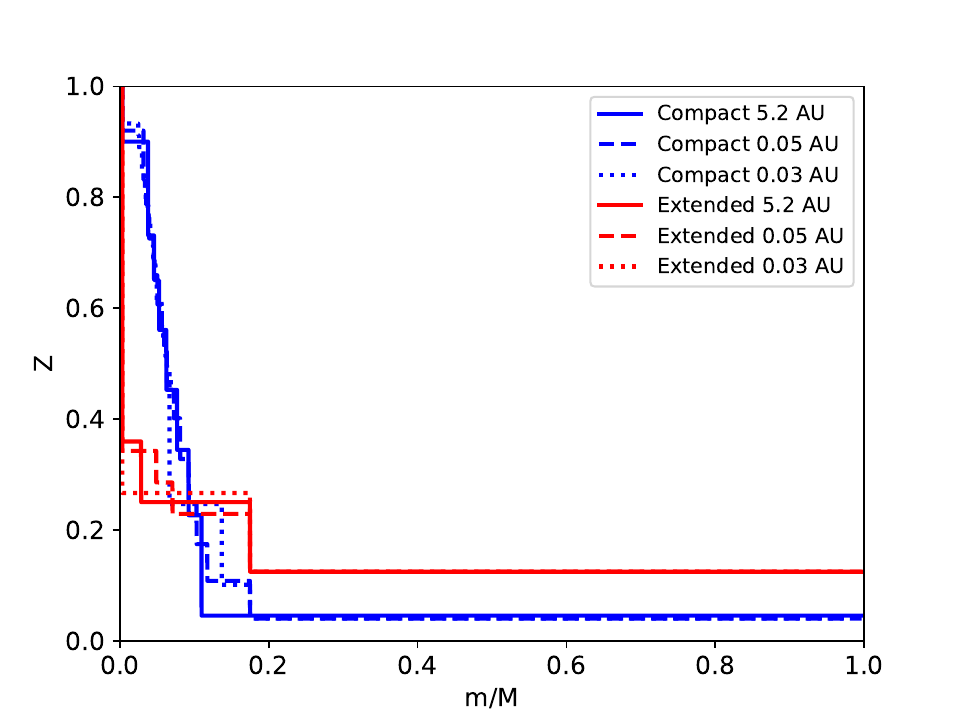}}
        \caption{Final composition of the compact and extended initial compositional structures for orbital distances of 0.05 and 0.03 AU. The standard case at 5.2 AU is shown for comparison. For the hot Jupiters, the bloating efficiency is in both cases 0.5\%}
    \label{Fig:Bloating_mix}
\end{figure}

The results for the model incorporating internal heating for a Jupiter-like planet at 0.05 and 0.03 AU is shown in Fig. \ref{Fig:Bloating_mix}. We find that the effect the additional internal heating has on the final structure is small, with the dilute core being retained in all cases. For the compact initial structure we find that the number of steps roughly doubles for the closer-in cases compared to the base case, with the extent of the dilute core also increases. For the extended structure the extent of the dilute core does not change at all, but for this structure the number of steps increases from 2 to 3 for the 0.05 AU case, while the number of steps reduces to only 1 for the 0.03 AU case. 

\begin{figure}
    \resizebox{\hsize}{!}{\includegraphics{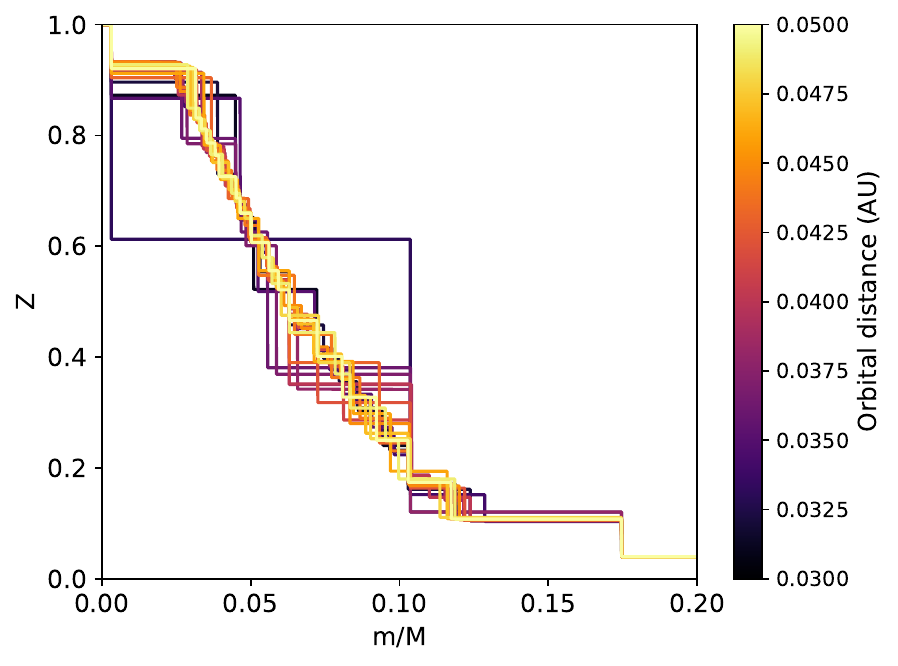}}
        \caption{Final composition of the compact initial compositional structure for orbital distances of 0.03 to 0.05 AU in 0.001 AU steps.}
    \label{Fig:Bloating_spread}
\end{figure}

Further analysis based on more orbital distances is shown in Fig. \ref{Fig:Bloating_spread}. This figure clearly illustrates the complexity of predicting the final structure based on the initial orbital distance. Each case leads to a unique final structure, although we can observe one clear trend. We observe that for distances between 0.04 and 0.05 AU mixing is suppressed compared to the same planet at 5.2 AU, while for closer-in planets mixing is enhanced. This effect can mainly be seen in the number of steps that are present, especially clear between 5 and 10\% of the mass. The extent of the dilute core does not change, since this is already limited by the initial compositional structure. The suppression of convective mixing is caused by the decrease in intrinsic luminosity due to the lower efficiency of cooling for hot Jupiters. For even closer-in hot Jupiters, the intrinsic luminosity increases to values higher than in our standard case, leading to more efficient convective mixing. This is the result of a large part of the envelope being hotter at the start of our simulation, while the dilute core itself is cooler at the same entropy. This leads to more efficient cooling early on, overcoming the overall effect of less efficient cooling for hot Jupiters. We are cautious to draw any conclusions from these results, since hot Jupiters are not expected to form in situ \citep{Lin+1996_HJMigration} and convective mixing would need to be suppressed through other means during migration for our results to hold true. For the extended initial structure we observe a similar pattern, with more steps being formed at smaller orbital distances, up to a point where this behaviour reverses, due to the effect described above. The main difference is in the extent of the dilute core, which does vary a little bit for this initial composition, which points to the importance of the initial compositional structure in determining these effects. The location of the steps and how they should be interpreted is further discussed in Sect. \ref{sec:steps}. 

The results we find are dependent on the assumed depth at which the energy is deposited and on the heating efficiency, which depend on the assumed heating mechanism. As mentioned before we deposit the energy at pressures derived for Ohmic heating by \citet{Ginzburg2016_OhmicHeating}, with a heating efficiency of 0.5\% that we found to best reproduce observed radii. We find that depositing the energy at lower pressures does not affect the results. For this reason we believe our results still hold for the advection of potential temperature, since this mechanism is expected to occur at lower pressures \citep{Tremblin+2017_AdvecPotTemp}. For other stellar heating mechanisms it is hard to determine the pressure at which the energy is deposited, since the efficiency of suppression of planetary cooling is just as efficient for heating at 10$^4$ bar as for heating at the centre of the planet. For thermal tides the depth of heat deposition remains uncertain \citep{Gu+2019_ThermalTides}. For most of these works a homogeneous envelope is assumed, but we do not believe this to have a significant impact, since the envelopes in our study are mostly homogeneous at pressures lower than 10$^6$ bar, where the energy is deposited. In extreme cases close to 30\% of the envelope mass is at pressures lower than 10$^6$ bar, but we find that for each of our models this region either starts homogeneous or becomes homogeneous during the early stages of evolution.

These results suggest that we should expect slightly more and larger dilute cores for hot Jupiters than for giant planets further out, provided they are formed in situ and are not too close to the star, although this is of course also dependent on the stellar type.

\subsection{Value of the mixing length parameter} \label{sec:alpha_MLT}
\begin{figure}
    \resizebox{\hsize}{!}{\includegraphics{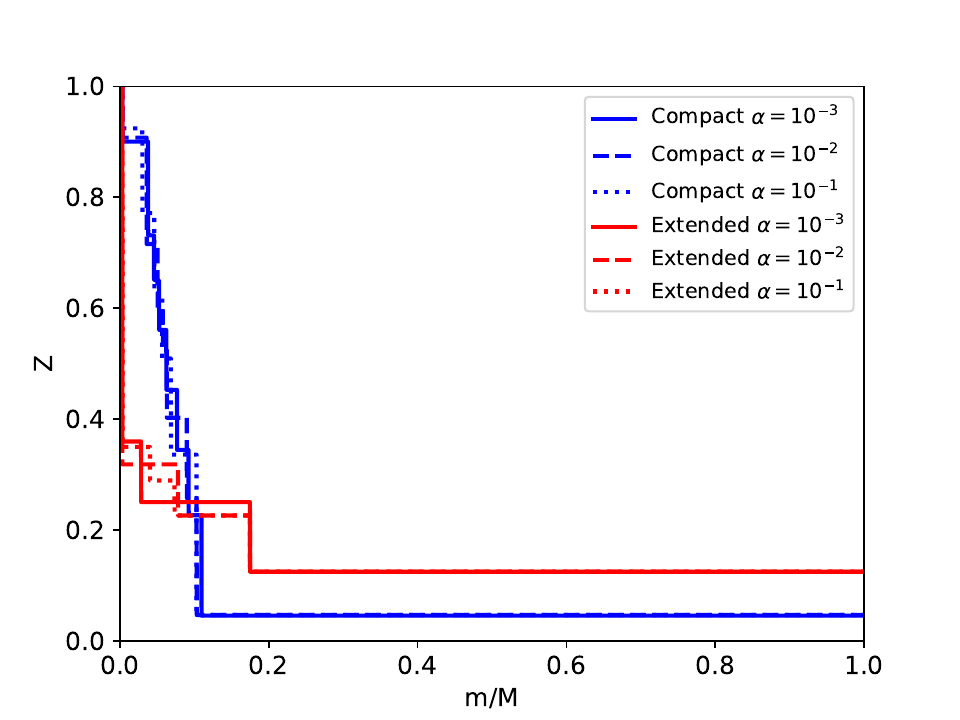}}
        \caption{Final composition of the compact and extended initial compositional structures for mixing length parameter values  of 10$^{-2}$ and 10$^{-1}$. The standard case of $\alpha=10^{-3}$ is shown for comparison.}
    \label{Fig:Alpha}
\end{figure}

The value of the mixing length parameter $\alpha$ for giant planets is unknown and directly determines the timescale for mixing within a convective zone. The effect of the value of $\alpha$ in our model is shown in Fig. \ref{Fig:Alpha}. The effect of the value of the mixing length parameter is minimal. The extent of the dilute core does not change for both the compact and extended initial compositions. Slight differences in the number of steps can be observed, but even the biggest of these, the extended model with $\alpha=10^{-2}$ and $\alpha=10^{-1}$ having three steps instead of two, does not have a large effect on the overall composition. This result is in line with the results of \citet{Vazan+2015_Convection}, who found that the value of $\alpha$ had very little impact on the final structure unless it was reduced to a value of 10$^{-6}$. 

\begin{figure}
    \resizebox{\hsize}{!}{\includegraphics{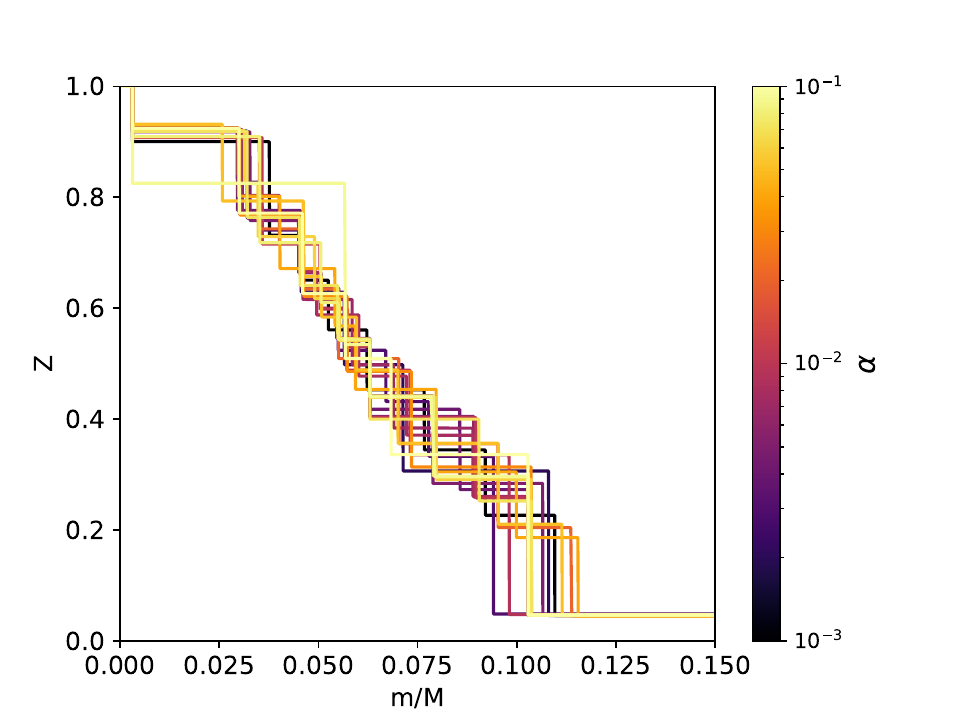}}
        \caption{Final composition of the compact initial compositional structure for values of $\alpha$ ranging from 10$^{-3}$ to 10$^{-1}$.}
    \label{Fig:Alpha_spread}
\end{figure}

To see whether any of the results described follow a trend, we looked at more values of the mixing length parameter, shown in Fig. \ref{Fig:Alpha_spread} for the compact initial structure. Similar to what we found for the orbital distance, and in line with the results from Fig. \ref{Fig:Alpha}, no strong relation can be found between the value of $\alpha$ and the final compositional structure of the planet. For the extended initial structure we also do not find any larger trends between the value of the mixing length parameter and the final compositional structure. The interpretation of the exact location of the steps and the cause of the differences between each final structure are discussed in Sect. \ref{sec:steps}.

\subsection{Semi-convection} \label{sec:semi-con}
\begin{figure}
    \resizebox{\hsize}{!}{\includegraphics{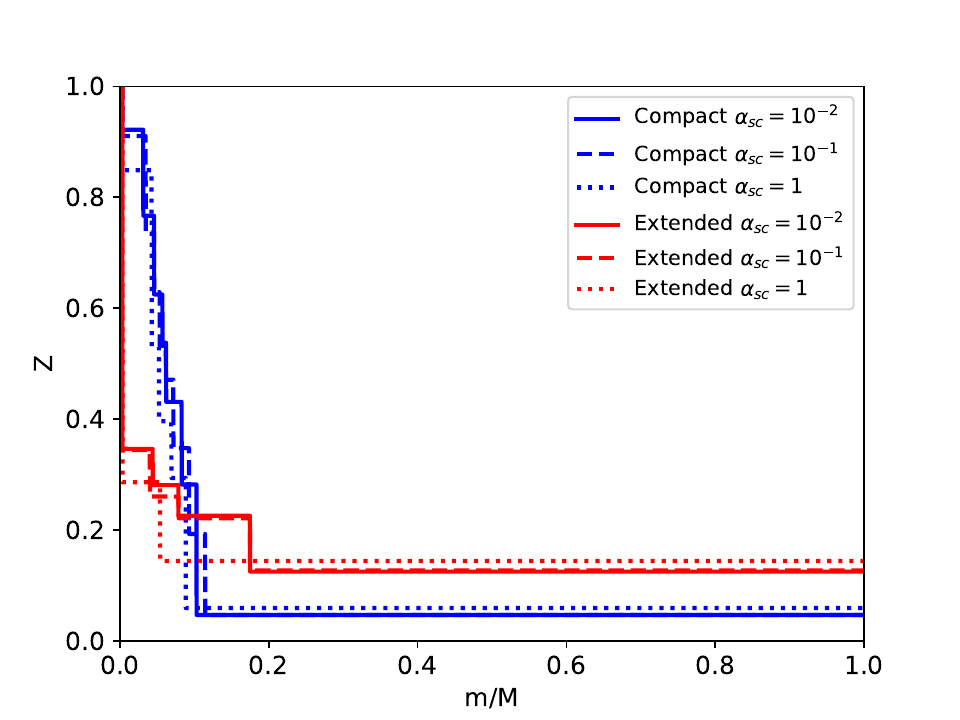}}
        \caption{Final composition of the compact and extended initial compositional structures for semi-convection parameter values, $\alpha_{sc}$, of 10$^{-2}$, 10$^{-1}$, and 1.}
    \label{Fig:Alpha_sc}
\end{figure}

Figure \ref{Fig:Alpha_sc} shows final compositions for different values of the semi-convection parameter $\alpha_{sc}$. For the compact initial structure we find that the dilute core extends slightly further out for a value of $\alpha_{sc}$ of 10$^{-1}$ than for $\alpha_{sc}$=10$^{-2}$ or $\alpha_{sc}=1$, although the difference is not very large. For $\alpha_{sc}=1$ the number of steps reduces to four, whereas there were seven without semi-convection. We also observe a decrease in the metallicity for large parts of the dilute core, which leads to an increase in metallicity in the outer envelope. For the extended initial structure the impact is stronger. Between $\alpha_{sc}$=10$^{-2}$ and $\alpha_{sc}$=10$^{-1}$ the difference is still small, with only the location of the inner steps changing slightly. With $\alpha_{sc}$=1 the dilute core extends to a much lower mass, also leading to an increase in high-Z material in the outer envelope. For both initial structures we find that strong semi-convection is necessary to affect the dilute core consistently. We also find that even with $\alpha_{sc}$=1, dilute cores are able to persist throughout the lifetime of the planet, although this does lead to a decrease in the size of the dilute core and an increase in the metallicity in the envelope outside the dilute core. We find that only for the Jupiter-like initial composition the dilute is completely mixed for values of $\alpha_{sc}\gtrsim10^{-2}$. \citet{Mueller+2020_Jupiter} also find that dilute cores can persist with strong semi-convection, but they do not find any differences in the extent of the dilute core or in the metallicity outside the dilute core. We find that the effect of the decreased temperature gradient due to semi-convection is small. We find that the difference in temperature is largest at the core-envelope boundary at the start of the simulation, with the temperature being $\sim$30 K cooler for $\alpha_{sc}$=1 compared to the standard case. As the planet evolves the higher mixing efficiency due to semi-convection plays a significantly larger role in cooling down the planet. \citet{Leconte+2012_semiconvection} suggest that the value of $\alpha_{sc}$ would have to lie somewhere between $10^{-9}-10^{-6}$ and $10^{-4}-10^{-2}$ for the solar system giants to be semi-convective on a large scale. This would severely limit the impact semi-convection could have on the compositional structure.

\subsection{Opacity}
\begin{figure}
    \resizebox{\hsize}{!}{\includegraphics{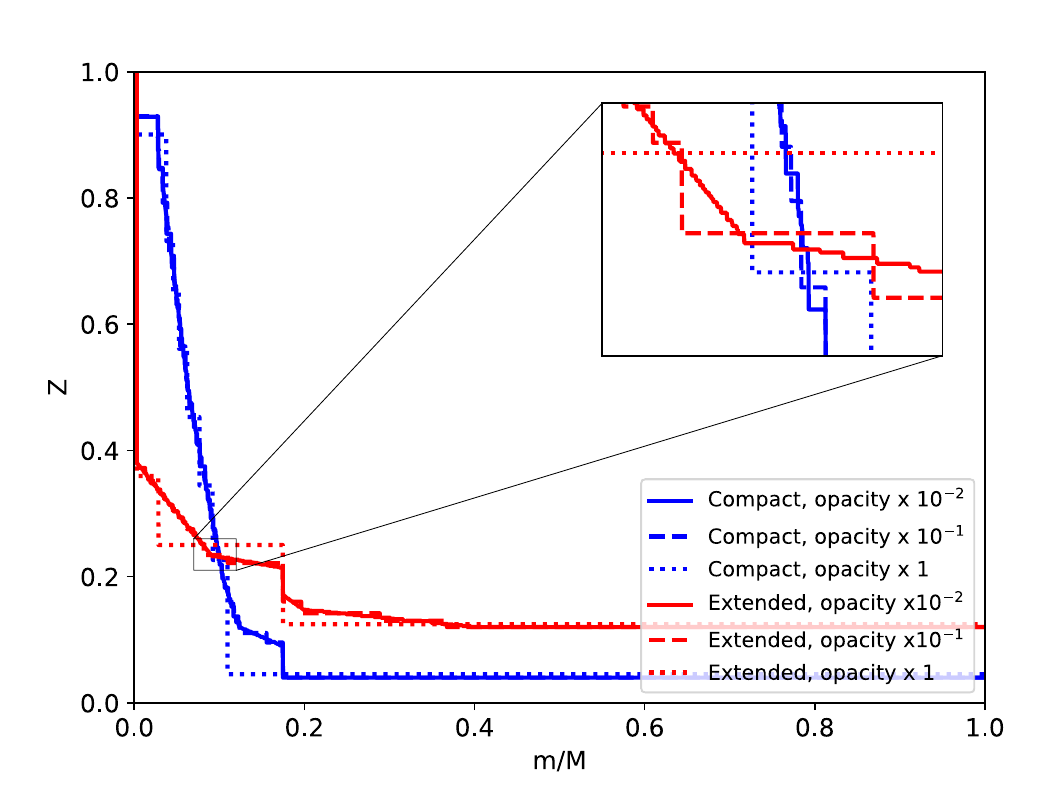}}
        \caption{Final composition of the compact and extended initial compositional structures for conductive opacities decreased by a factor of 10 and 100. The standard case using the opacity as given is shown for comparison. The inset shows a closer look at the differences in the step structure.}
    \label{Fig:Opacity}
\end{figure}

The opacity plays an important role in the cooling of planets. It also has a strong influence on the radiative-conductive gradient and is thus an important factor in determining when convection is inhibited. In nearly the entire envelope, the opacity is governed by the conductive opacity, for which we used the values given by \citet{Cassisi2007_condopacities}. \citet{Podolak+2019_Conductivity} find that the values given by \citet{Potekhin1999_ConductiveOpacity}, which agree within an order of magnitude with those of \citet{Cassisi2007_condopacities}, are up to two orders lower than those found in {ab initio} calculations using density functional theory by \citet{French+2017_ElectronicConductivity} and \citet{French2019_ThermalConductivity}. The radiative opacity also has an effect on the cooling rate, and thus indirectly on the mixing process. Here we focused on the conductive opacity, due to its larger role in the mixing process and the large uncertainty regarding its value. To analyse the effect of the conductive opacity, we multiplied its values by 0.01, 0.1, and 10. We find that increasing the opacity by a factor of 10 leads to complete mixing for all the initial compositions we looked at. This is not very surprising, as the radiative-conductive gradient scales linearly with the opacity, while the compositional gradient is barely affected. Lowering the opacity leads to an increase in the number of compositional steps, as well as an increase in the extent of the dilute core. This effect is shown in Fig. \ref{Fig:Opacity}. For the compact structure we find that lowering the opacity leads to an increase in the number of steps and to the dilute core extending further out, to 17\% of the mass, compared to 11\% for the standard case. For the lowered opacity the extent of the dilute core is limited by how far the dilute core extends initially. For the extended structure we see similar results. Lowering the opacity to 0.1 of the original value leads to the dilute core extending significantly further out. Lowering the opacity by another factor of 10 only has a very small impact on the compositional structure. The inset in Fig. \ref{Fig:Opacity} shows the effect the decrease in opacity has on the number of steps in the structure, which increases when the opacity is decreased. We do not expect the actual opacity to be 0.01 of the value given by \citet{Cassisi2007_condopacities}, since this leads to rapid cooling of the interior and high present-day luminosities of $\sim$1.4 L$_J$ compared to $\sim$0.6 L$_J$ for the standard case. Interestingly using 0.1 or 10 times the standard opacity barely affects the present-day luminosity. It is interesting to note that our results differ from those of \citet{Mueller+2020_Jupiter}, who find that the effect of the opacity on the final composition is small. They do find a similar relation, with stairs becoming less stable at higher opacities and the dilute core extending to higher masses at lower opacities, but the strength of these effects is much smaller than what we find. It is unclear what the exact cause of this difference is, but it is likely caused by general differences between the models and not by the opacity directly.

\subsection{Number of mesh points} \label{sec:gridpoints}
\begin{figure}
    \resizebox{\hsize}{!}{\includegraphics{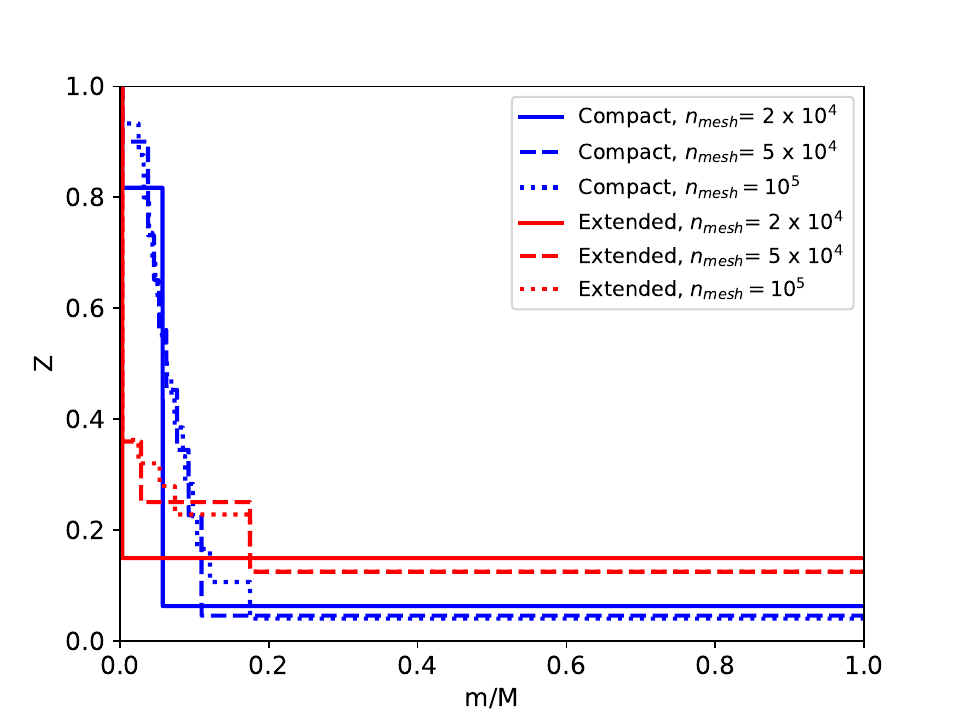}}
        \caption{Final composition of the compact and extended initial compositional structures for $2 \times 10^4$ and 10$^5$ mesh points. The standard case with $5 \times 10^4$ mesh points is shown for comparison.}
    \label{Fig:ngrid}
\end{figure}

As mentioned before, we used a number of mesh points of $5 \times 10^4$. This value is significantly higher than that used in \citet{Vazan+2015_Convection} and \citet{Vazan+2018_Jupiter}, where values of 150 and 500 were used. Direct comparison is complicated by the fact that their mesh does not consist of equal mass-steps, but follows the requirement of constant increments in 
\begin{equation}
    f=(m/M)^{2/3}+c_1 X_H -c_2 ln p -c_3 ln \frac{T}{T+c_4} 
\end{equation}
given by the model \citep{Kovetz2009_Vazancode}. \citet{Mueller+2020_Jupiter} does not use a fixed mesh, but follows the mesh given by MESA \citep{Paxton2011_MESA1,Paxton2013_MESA2}, where cells are split or merged automatically to ensure numerical accuracy and efficiency. Each method has the same problem, in that the size of the mesh determines the maximum compositional gradient for a given change in metallicity, which directly determines whether mixing can occur or not. This can be understood by analysing Eq. \ref{eq:nablaX}. $\frac{\partial ln T(\rho,p,X)}{\partial X_J}$ is given by the equations of state and the local pressure and density. The second part is $\frac{dX_j}{dlnp}$, where dX$_j$ is the change in metallicity, which is limited by the difference between two cells, and $dlnp$ is limited by cell size, which follows from the number of mesh points that are used in the model. For a smooth gradient these two factors scale linearly with the cell size, meaning that the number of mesh points does not affect whether mixing occurs. In the presence of steps, however, dX$_j$ is limited by the size of the step, while $dlnp$ still scales with the cell size, meaning that the cell size dictates the size of steps necessary to inhibit convection and whether steps can inhibit convection at all. \citet{Vazan+2015_Convection} also discuss this issue and point out the effect this has on the number of steps, but we find the effect to be more significant in our study, where the initial luminosity is greater. In our case, the number of mesh points affects whether a dilute core persists and it has an impact on the extent of the dilute core. This effect is shown in Fig. \ref{Fig:ngrid}. We find that when using $2 \times 10^4$ mesh points only the compact initial structure does not become fully mixed, forming a single step and only extending to $\sim$6\% of the mass. Increasing the mesh size to 10$^5$ points roughly doubles the number of steps in comparison to the standard case and leads to a dilute core that extends to $\sim$17\% of the mass. For the extended case the change is less severe, only leading to two extra steps being formed and the dilute core extending to the same mass as with $5 \times 10^4$ mesh points. 

It is unclear what number of mesh points is most realistic, or how these mesh points should be distributed throughout the planet. \citet{Leconte+2012_semiconvection} find that Jupiter cannot consist of more than $3 \times 10^4$ convective/diffusive layers based on observational constraints. Each of these layers are composed of a larger convective cell and a small diffusive interface with a compositional gradient. Accurately modelling such a composition would require several times the number of layers in mesh point, although it is unclear what the effect of larger convective cells and larger compositional gradients would be. One can also argue that the size of the mesh should be significantly larger than the overshooting length scale, at least as long as overshooting is not included in the model. \citet{Hindman2023_RotationEntrainment} find that the overshooting length for a Jupiter-like planet is $\sim3.6 \times 10^5$ cm shortly after formation. Simply dividing the radius of the planet by this length just after formation would give $\sim3 \times 10^4$ mesh points. This is less than the value of $5 \times 10^4$ mesh points we used, which begs the question as to whether this is a reasonable choice. However, the overshooting length decreases by a factor of $\sim$10 when using our initial luminosity of 10$^3$ L$_J$, which would correspond to $\sim3 \times 10^5$ mesh points. To solve this problem, the overshooting must be accurately modelled, which we have not attempted in this work.

\subsection{Initial luminosity}  \label{sec:InitLum}
\begin{figure}
    \resizebox{\hsize}{!}{\includegraphics{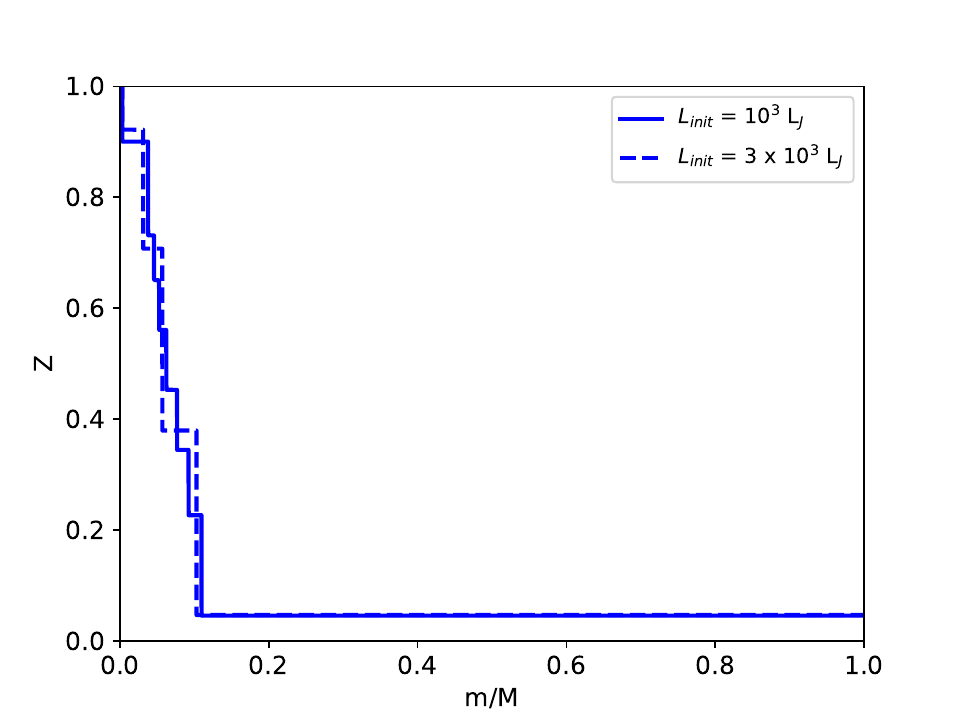}}
        \caption{Final composition of the compact initial compositional structures for an initial luminosity of 3 $\times$ 10$^3$ L$_J$. The standard case with an initial luminosity of 10$^3$ L$_J$ is shown for comparison.}
    \label{Fig:Luminosity}
\end{figure}

Similar to the number of mesh points, the initial luminosity plays a significant role in determining whether mixing occurs. Whereas the number of mesh points determines the maximum of the compositional gradient, the initial luminosity plays a large role in determining the radiative-conductive gradient, with the two scaling roughly linearly. This is reflected in our results for models with higher initial luminosities. The results for an increase in the luminosity are shown in Fig. \ref{Fig:Luminosity}. We find that with an initial luminosity of $3 \times 10^3$ L$_J$, only the compact and metal-rich initial composition retain their dilute core, with only one step remaining. Increasing the luminosity to 10$^4$ L$_J$ leads to the envelope becoming fully mixed for these compositions as well. This strong relation between the degree of mixing and the luminosity was also found by \citet{Knierim+2024_ConvectiveMixing}, although they varied the entropy instead of the luminosity.  

It is important to note that while each initial composition starts at the same initial luminosity, the luminosity increases at the start of the evolution due to the mixing process. This is most likely also a result of the differences between the formation models used to get to these structures and our evolution model. The maximum luminosity reached can be interpreted as the `true' initial luminosity. For an initial luminosity of $10^3$ L$_J$, the maximum luminosity reached for the compact, extended, metal-rich, and Jupiter-like initial composition are $1.3 \times 10^3$ L$_J$, $1.1 \times 10^3$ L$_J$, $1.8 \times 10^3$ L$_J$, and $1.5 \times 10^3$ L$_J$, respectively. For an initial luminosity of $3 \times 10^3$ L$_J$ the maximum luminosities are $3.6 \times 10^3$ L$_J$, $3.1 \times 10^3$ L$_J$, $5.7 \times 10^3$ L$_J$, and $4.6 \times 10^3$ L$_J$. Redoing the simulations using the final composition as the initial structure does not lead to an increase in luminosity, but reverts to the expected result of a slow continuous decrease in luminosity. This supports the interpretation that the luminosity increase is a direct result of the mixing process. 

Overall the initial luminosity is shown to be a significant factor in whether dilute cores are able to persist. While the initial increase in luminosity is significant, it still seems unlikely for any dilute core to persist at luminosities much higher than $3 \times 10^3$ L$_J$. This suggests that dilute cores would not be able to persist in planets formed through a hot-start scenario \citep{Burrows1997_HotStart,Baraffe2003_HotStart}, as such planets form with significantly higher initial luminosities, although their luminosities could be lower if large radiative-conductive zones emerge \citep{Berardo2017_AccretionShock}. \citet{Marley2007_coldstart} predicts that 1 M$_J$ planets formed through a cold-start scenario are expected to have initial luminosities roughly two times higher than $3 \times 10^3$ L$_J$, while \citet{Mordasini2013_Luminosity} finds luminosities of $\sim2 \times 10^4$ L$_J$ to $\sim6 \times 10^4$ L$_J$ for cold accretion depending on core mass. It is possible that some planets formed in a cold-start scenario could be capable of retaining a dilute core in our model, as long as a large enough compositional step is present, but under the assumptions of our model it is unlikely that this would be the case for a large percentage of giant planets.

\section{Discussion} \label{Sec:Limits}
\subsection{Interpretation of steps and numerical behaviour} \label{sec:steps}
One of the largest difficulties with modelling convective mixing is interpreting the appearance of steps and relating this to observational quantities. \citet{Vazan+2018_Jupiter}  discussed the appearance of steps or stairs and concludes that while their existence is most likely physical, it is difficult to determine their exact location and size due to the dependence on the number of mesh points. As discussed in Sect. \ref{sec:gridpoints}, determining what number of mesh points to use remains a difficult problem, meaning that the problem with steps will not be solved easily. Still there is a lot be understood from them, as long as we do not interpret them as exact predictions of the compositional structure, but rather as regions of compositional gradients. Both Figs. \ref{Fig:Bloating_spread} and \ref{Fig:Alpha_spread} illustrate this point. While the general structure remains similar for different values of the orbital distance or mixing length parameter, the exact location, the size, and the number of steps changes, with no consistent pattern from one value to the next. We observe this effect across many parameters, both parameters for which we also observe and expect a relation with the final structure, such as the orbital distance, and parameters for which be do not observe a consistent relation with the final structure, such as the mixing length parameter $\alpha$. So while the specific compositional structures found with our model do not reflect the exact structure, they can be useful for finding and interpreting general trends related to convective mixing.   

\subsection{Delayed convection} \label{sec:delayedconv}
As mentioned in Sect. \ref{Sec:Evo_model}, we find that for some of the initial compositions convective mixing happens immediately and mixes the entire envelope. This is due to only shallow compositional gradients being present with no steps large enough to inhibit convection. Once the entire planet becomes convective, mixing will lead to complete homogeneity. For the extended and metal-rich initial compositions we find that convection does not need to be inhibited to retain a dilute core, although the extent of the dilute core is less in this case. The fact that the dilute core can be retained is likely caused by there already being a large enough step in the compositional gradient in the initial composition to prevent convection. This is supported by the fact that the step in the final composition is at the same mass as the step in the initial composition. For this metal-rich composition we find that the inhibition of convection does not have a significant effect on the other planetary properties throughout the evolution. 

It could be argued that limiting convection initially could be a reasonable explanation based on the assumption of hot (high entropy) accretion, which can prevent convection for $10^6$ years after accretion has stopped for a 1 M$_J$ planet \citep{Berardo2017_AccretionShock}. \citet{Mueller+2020_Jupiter} incorporate such radiatively inefficient accretion shocks in their hot-start formation models, which might explain why we need to inhibit convection initially to get similar results. Inefficient accretion shocks however, are not present for cold-start scenarios, so discrepancies for these scenarios cannot be explained by this. It seems that the overall discrepancies between the formation model used by \citet{Mueller+2020_Jupiter} and our evolution model are the main cause of this difference. Independent of formation pathways, the radiative-conductive gradient increases relatively slowly due to the increase in luminosity when compared to the mixing timescale. Based on this one can expect small steps to form early on during formation, which then merge into larger steps as the radiative-conductive gradient and luminosity continue to increase. 

\subsection{High internal temperatures} \label{sec:hightemps}
We find high inner envelope temperatures of up to $1.8 \times 10^5$ K, which is significantly higher than the initial internal temperatures of $\sim5 \times 10^4$ K found in previous works \citep{Vazan+2015_Convection,Vazan+2018_Jupiter,Mueller+2020_Jupiter} for similar structures. The difference in temperature can be explained by \citet{Mueller+2020_Jupiter} using temperature gradients based on the work of \citet{Kippenhahn2012_MLTequations}, which do not incorporate compositional gradients in the determination of the temperature gradients. This can lead to significant differences in the presence of large compositional gradients. Discrepancies with \citet{Vazan+2018_Jupiter} can be explained by the large difference in initial luminosities, with our higher initial luminosity leading to higher internal temperatures. It is also relevant to note that the Jupiter-like initial composition, which is based on the work of \citet{Vazan+2018_Jupiter}, has the lowest internal temperature out of the initial compositions we used, most likely due to the compositional gradients being smaller than for the other structures. The main model of \citet{Vazan+2015_Convection} with a continuous Z-gradient has an internal temperature of only $\sim 5 \times 10^4$ K at early times during the evolution, but several other with more extreme compositional gradients still have an internal temperature of $\sim 8 \times 10^4$ K after 10 Gyr, with the planets likely having had an even higher temperature at earlier times. So while the internal temperatures we reach are quite high, it does not seem like there are large inconsistencies with previous works. 

\begin{figure}
    \resizebox{\hsize}{!}{\includegraphics{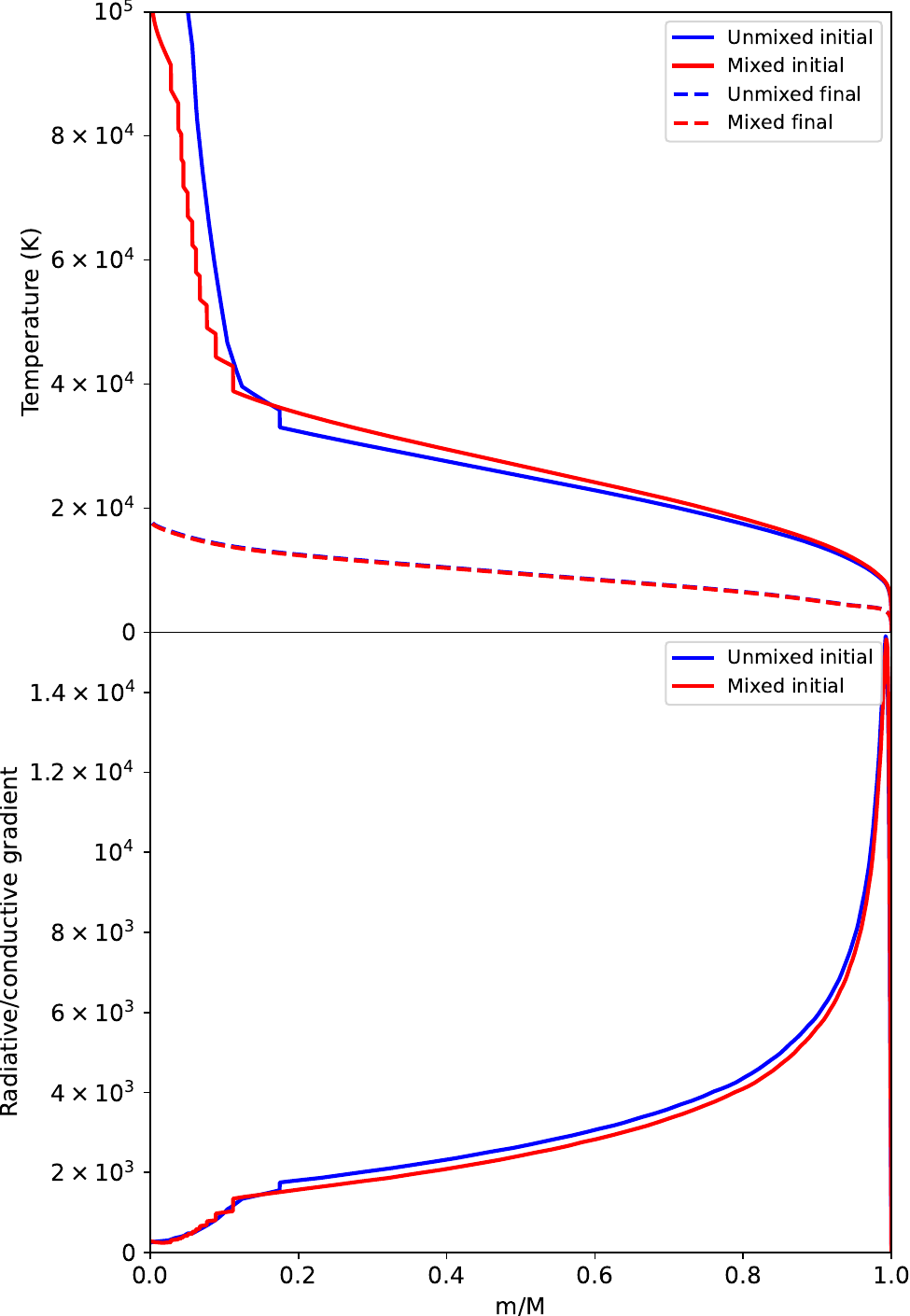}}
        \caption{Temperature (top) and radiative-conductive gradient (bottom) for the compact compositional structure in its original state (unmixed) or starting with its final compositional structure (mixed). The temperature is shown at both the start of the simulation and after 4.5 Gyr, while the radiative-conductive gradient is only shown at the start. The final temperature structures are very similar, making it hard to distinguish between them.}
    \label{Fig:Mixed}
\end{figure}

To fully analyse the effect of the high internal temperatures, we ran tests using the final compositional structures of our standard models as the initial structure, while maintaining the same initial luminosity. This method makes the inconsistencies between the formation model of \citet{Mueller+2020_Jupiter} and our evolution model less relevant, as the mixing has already occurred. Figure \ref{Fig:Mixed} shows a comparison of the temperature in the original compact composition and the already mixed version based on the standard case. Since the compact composition had the highest initial internal temperature it is the ideal candidate for illustrating the differences. We find that the mixed structures show significantly lower internal temperatures, due to less of the planet being affected by compositional gradients. In the steps, where the gradients are present, the temperature increase is significantly less than in the original compositions where the compositional gradients are relatively smooth. Even though the temperatures at the start of the simulation are very different, the temperature profiles converge after mixing occurs during the 4.5 Gyr of evolution, with the increased temperature gradient in the steps becoming smaller and the temperature structure becoming almost completely adiabatic. We find that the difference in the radiative-conductive gradient, which is given as in Eq. \ref{eq:radGrad}, is significantly smaller than the difference in temperature, although the radiative-conductive gradient in the case where the planet is already mixed can be up to 14\% smaller than in the unmixed state around $\sim$0.2 m/M for the compact structure. Interestingly, for the other initial structures the differences reduce to only a few percent at the relevant masses, most likely due to the much smaller differences in the temperatures between the mixed and unmixed state. We also find that the compositional gradient differs between the initially unmixed and mixed models caused by the differences in the temperature and pressure, but this difference is minimal and also converges over time. 

To summarise, the high internal temperatures we find do have an effect on convective mixing, although this effect is smaller than one might naively expect from the high temperatures. If more mixing occurred during formation or the internal temperature were lowered in another way, this would lead to a decrease in convective mixing. The radiative-conductive gradient becomes smaller at lower temperatures, which aids in inhibiting convective mixing. In most cases the effect is limited to only a few percent, but in cases where there is a large compositional gradient near the core, such as for the compact initial structure, this effect can be larger.

\subsection{Luminosity distribution in the envelope}\label{sec:luminosity_simp}
\begin{figure}
    \resizebox{\hsize}{!}{\includegraphics{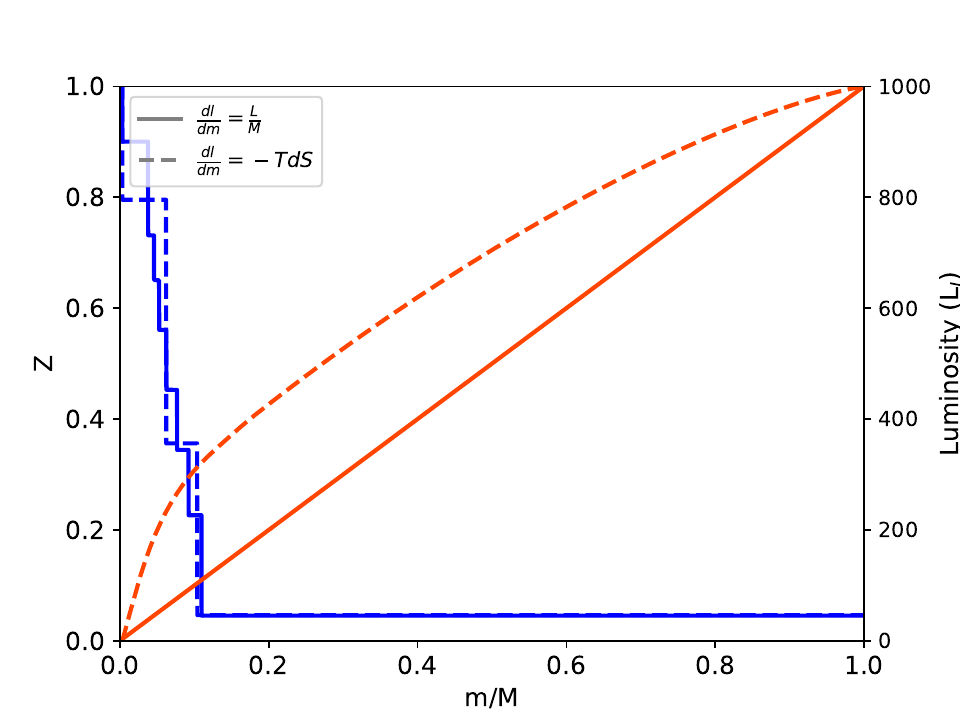}}
        \caption{Final composition of the compact structure, comparing two simplifications of determining the luminosity and an illustration of the difference in luminosities at the start of the evolution. The solid line shows the composition with the luminosity scaling linearly with the mass, whereas the dashed line shows the composition with luminosity scaling as T$\frac{dS}{dt}$ with the mass, where $\frac{dS}{dt}$ is taken to be constant.}
    \label{Fig:Lumishape}
\end{figure}

As mentioned in Sect. \ref{Sec:Method}, we used the simplified method of $\frac{dl}{dm}=\frac{L}{M}$ for the distribution of the luminosity in the envelope. Our previous work used the simplification $\frac{dl}{dm}=0$ \citep{Mordasini+2012_Completo}, which is adequate under the assumption that the interior is fully convective, but not sufficient for our goals. Realistically we would use the full equation $\frac{dl}{dm}=-T\frac{\partial S}{\partial t}$, but this is difficult to model, due to it being hard to self-consistently determine the change in entropy locally, while using our global energy-conservation approach to estimate the total luminosity. This is made more complicated by the fact that convective mixing affects the entropy as well. To explore the consequences of our simplification, we also tested an alternative simplifying method with the equation $\frac{dl}{dm}=-T \frac{dS}{dt}$, where $\frac{dS}{dt}$ is a constant throughout the structure such that the boundary conditions are satisfied. We find that this method is numerically  significantly less stable than the $\frac{dl}{dm}=\frac{L}{M}$ method, although we still manage to reach convergence when using appropriate parameters. This method leads to larger luminosity near the core, due to the high temperatures there. The difference in luminosity can be up to a factor of 3 around 0.1 to 0.2 m/M, an especially relevant mass range as this is the usual extent of the dilute core. An example of this difference is shown in Fig. \ref{Fig:Lumishape} for the compact initial structure. This difference also extends to the radiative-conductive gradient, which is also found to differ up to a factor of 3 in this range. The effect on the final compositional structure can also be seen in Fig. \ref{Fig:Lumishape}. The number of steps decreases from 7 to 2 while the dilute core extends to the same mass in both cases. For the other initial structures the effect is significantly weaker, with the luminosity and radiative-conductive gradient differing by less than a factor of 2 in the 0.1 to 0.2 m/M range, although a decrease in the number of steps still occurs in these cases. Since the luminosity structure is dependent on the temperature for the $\frac{dl}{dm}=-T \frac{dS}{dt}$ implementation, we also ran tests using the already mixed compositions, which have lower internal temperatures near the core. We find that this reduces the maximum difference to a factor of 1.8, significantly less than the factor of 3 we found with the high internal temperatures. In line with \citet{Mordasini+2012_Completo} we find that the evolution is not significantly affected by the change to the luminosity, besides what is expected from the changes in the compositional structure. It is important to note that even the $\frac{dl}{dm}=-T \frac{dS}{dt}$ luminosity distribution is still a simplification where the local changes in entropy are not actually considered, due to the limitations of the code. \citet{Knierim+2024_ConvectiveMixing} found that entropy gradients play a significant role in the mixing process and that the primordial entropy profile, and not just the average primordial entropy, can significantly impact the final heavy element distribution.
The fact that the luminosity distribution using $\frac{dl}{dm}=-T \frac{dS}{dt}$ leads to different results by a factor of a few compared to our standard $\frac{dl}{dm}=\frac{L}{M}$ luminosity distribution and the role the high internal temperature might play, combined with the importance of accurately tracking the entropy, mean that it will be important to calculate the luminosity and entropy more accurately in future studies and to be mindful of the limits of the current implementation. 

\subsection{Additional limitations}
So far in this section we have addressed the major limitations of our work. Here we discuss some of the additional effects that have been suggested to affect the mixing process.

The condensation of heavy species has been suggested to possibly inhibit convection in the atmosphere of giant planets \citep{Guillot1995_Condensation,Leconte+2017_Condensation}. This could significantly reduce the cooling rates of these planets, which would lower the expected amount of convective mixing. It is still unclear to what extent this process is present in the solar system gas giants and how big the effect might be for extrasolar giants.

Recent works have shown that planetary rotation can significantly reduce the effectiveness convective mixing in giant planets \citep{Fuentes2023_Rotation,Hindman2023_RotationEntrainment,Fuentes+2024_RotationSemicon}. \citet{Fuentes2023_Rotation} find that rotation reduces the convective velocity by a factor of 6 for Jupiter, which in turn reduces the kinetic energy flux available for mixing by a factor of $\sim$200. \citet{Hindman2023_RotationEntrainment} find that planetary rotation could prevent a planet from becoming fully convective. \citet{Fuentes+2024_RotationSemicon} find that semi-convective staircases in Jupiter could be conserved for 10$^{11}$ years compared to 10$^9$ years for a non-rotating planet. Since these effects are not incorporated in our model, they could also significantly lower the amount of convective mixing.

In our study we only used the equations of state of \citet{Chabrier+2021_HHeEOS} for a mixture of hydrogen and helium and the equations of state of \citet{Haldemann+2020_H2OEOS} for water. When using water as our heavy element material we significantly underestimate the mean molecular weight for a planet whose heavy element mass consists of SiO$_2$ or other heavy molecules. Theoretically an underestimation of the mean molecular weight overestimates the amount of mixing in the planet, due to the compositional gradient being strongly related to the mean molecular weight gradient. Previous studies \citep{Vazan+2015_Convection,Vazan+2016_JupSaturnConv,Knierim+2024_ConvectiveMixing} have found significantly less mixed interiors when using SiO$_2$ as the heavy element material compared to water. \citet{Mueller+2020_Jupiter} surprisingly found that the structure was slightly  better mixed with SiO$_2$, although the extend of the dilute core did not change.

\section{Difficulty of retaining a dilute core} \label{Sec:retaincore}
For distant and close-in Jovian mass planets, we find a range of parameters for which we are able to retain a dilute core, but this parameter space is quite small, making it unlikely that dilute cores are retained in a large fraction of giant planets, under the assumptions and limitations of our model. We find that nearly all mixing occurs at early stages of evolution, requiring the formation of large compositional steps during formation to retain a dilute core. Steps will be unable to be retained once mixing occurs across the boundary, as this will cause the compositional gradient to decrease in size on much shorter timescales than it takes for the radiative-conductive gradient to decrease in size as the result of cooling. As a result of this, it is possible to define requirements for retaining a dilute core that must be satisfied, under the assumption that no major events occur during the evolution of the planet. The main requirements concern the steps in the composition and the initial luminosity. The limits of these values are dependent on what number of mesh points we used in our model, and as discussed in Sect. \ref{sec:gridpoints} it is hard to determine what number would be best to model the situation in gas giants. 

Assuming that the number of mesh points of 5 $\times$ 10$^4$ that we have used so far provides a good approximation, we need a compositional step of $\gtrsim$0.05 in Z to prevent mixing. The necessary size of the step also depends on the location, with it being harder for steps to persist at higher masses and being very unlikely for any steps to persist beyond 0.4 m/M, due to the increase in the radiative-conductive gradient. We also require an initial luminosity of $\lesssim$ 3 $\times$ 10$^3$ L$_J$, where we consider the maximum luminosity reached as the initial luminosity as discussed in Sect. \ref{sec:InitLum}. It might be possible to retain a dilute core at higher luminosities, but this would require larger steps in the composition. Similarly, a smaller compositional step could be enough to prevent mixing if the initial luminosity were lower. Unless the number of mesh points we used is significantly lower than necessary to model the actual process, we conclude that it is unlikely that dilute cores are retained for a large number of planets based on our current understanding of planet formation pathways. The formation and retention of dilute cores through other processes, such as giant impacts \citep{Liu+2019_GiantImpacts}, has not been considered here and requires further investigation.

\section{Summary} \label{Sec:summary}
In this work we modelled the evolution of giant planets, incorporating convective mixing. We analysed several initial compositions and determined the importance of parameters relevant to the mixing process. We compared our results to existing calculations of distant Jupiter analogues to benchmark our model, and we present new domain results for hot Jupiters. Our results, under the assumptions and limitations of our model, can be summarised as follows:
\begin{itemize}
    \item The initial composition has a large effect on the size and shape of the final dilute core. Large compositional gradients help create steps, while a large spread in the initial composition helps form dilute cores that extent farther out.
    \item Additional internal heating for hot Jupiters can decrease the cooling rate, leading to dilute cores being more easily retained compared to cold Jupiters, although the effect is small.
    \item For stronger additional internal heating, the envelope becomes relatively hot compared to the dilute core, causing dilute cores to become harder to retain.
    \item Effective semi-convection can alter the composition of the planet, but the effect is not strong enough to fully mix a planet where a large dilute core is present initially. 
    \item Dilute cores are not able to persist for initial luminosities much higher than $3 \times 10^3$ L$_J$ for a 1 M$_J$ planet, due to the relation between the luminosity and the radiative-conductive gradient.
    \item The number of mesh points is an important parameter in determining whether a dilute core is retained. It remains difficult to determine what number of mesh points is reasonable, and determining such a value accurately requires further research. 
\end{itemize}

Our model suggests it is unlikely that a large number of giant planets are able to retain an extended dilute core throughout their evolution assuming standard formation pathways, although this result remains dependent on the assumptions and limitations of our approach. Work remains to be done on improving evolution models and on understanding the mixing process in greater detail, especially near radiative-convective boundaries. JWST can also provide constraints for the prevalence and size of dilute cores via accurate measurements of the atmospheric metallicity combined with mass and radius data.

\begin{acknowledgements}
We thank the referee for useful comments and suggested additions to this paper. This work has been carried out within the framework of the National Centre of Competence in Research PlanetS supported by the Swiss National Science Foundation under grant 51NF40\_205606. The authors acknowledge the financial support of the SNSF.
\end{acknowledgements}

\bibliographystyle{aa}
\bibliography{bibliography.bib}

\end{document}